\documentclass[a4paper,conference]{IEEEtran}
\IEEEoverridecommandlockouts
\usepackage{cite}
\usepackage{amsmath,amssymb,amsfonts}
\usepackage{algorithmic}
\usepackage{graphicx}
\usepackage{textcomp}
\usepackage{xcolor}
\usepackage{subfig}
\usepackage{graphicx}
\usepackage{booktabs}
\usepackage{tabularx} 
\def\BibTeX{{\rm B\kern-.05em{\sc i\kern-.025em b}\kern-.08em
    T\kern-.1667em\lower.7ex\hbox{E}\kern-.125emX}}
\begin{document}

\title{CDMA and Non-Uniform Multiplexing: Dynamic Range in MIMO Radar Waveforms}
\author{%
\IEEEauthorblockN{%
Aitor Correas-Serrano{$^{\text{\#}1}$}, 
Nikita Petrov{$^{\text{*\&}2}$}, 
Maria Gonzalez-Huici{$^{\text{\#}3}$},
Christian Greiff{$^{\text{\#}4}$}, 
Alexander Yarovoy{$^{\text{*}5}$},
}

\IEEEauthorblockA{%
{$^\text{\#}$}Fraunhofer FHR, Germany\\
$^\text{*}$TU Delft, Netherlands\\
$^\text{\&}$NXP Semiconductors, Netherlands\\
\{$^1$Aitor.Correas, $^3$Maria.Gonzalez, $^4$Christian.Greiff\}@fhr.fraunhofer.de, \\
\{$^2$N.Petrov\}@nxp.com   \\
\{$^5$A.Yarovoy\}@tudelft.nl   \\
}% \IEEEauthorblockA Affils

}

\maketitle

\begin{abstract}
This paper presents a performance comparison of various MIMO radar multiplexing approaches where the increasing number of transmitters adversely affects the dynamic range of the resultant MIMO system. The investigated multiplexing techniques are code-division multiplexing in phase-coded frequency-modulated continuous wave (FMCW) and phase-modulated continuous wave (PMCW) radar. Additionally, random fast/slow time multiplexing in time and frequency is considered for frequency-modulated continuous wave and orthogonal frequency-division multiplexing (OFDM) radars. The comparative analysis is conducted through simulations, evaluating the scalability with the number of transmitters and addressing other pertinent implementation concerns. The findings provide insights into the trade-offs associated with each approach, aiding in selecting suitable MIMO radar multiplexing strategies for practical applications.
\end{abstract}


\section{Introduction}

As radar chips designed for low-cost civil applications support an increasing number of transmitters and receivers, conventional multiplexing approaches based on time-division multiplexing (TDM) and Doppler-division multiplexing (DDM) are becoming less attractive due to the significant reduction in maximum unambiguous velocity that can be estimated. Instead, code-based alternatives and random multiplexing in time and frequency can shift the MIMO trade-off into the dynamic range domain.

%\begin{figure*} 
%
%    \centering
%  \subfloat[\label{patt_a}]{%
%       \includegraphics[width=0.45\linewidth]{Figures/fmcw_ol}}
%       \qquad
%  \subfloat[\label{patt_b}]{%
%        \includegraphics[width=0.45\linewidth]{Figures/pmcw_ol}}    
%    \hfill 
%
%    \subfloat[\label{patt_c}]{%
%       \includegraphics[width=0.45\linewidth]{Figures/ofdm_ol}}
%       \qquad
%  \subfloat[\label{patt_d}]{%
%        \includegraphics[width=0.45\linewidth]{Figures/pc_fmcw_ol}}    
%    \hfill        
%  \caption{Schematic depiction of studied waveforms and multiplexing schemes: (a) and (c) show non-uniform time and time-frequency multiplexing for FMCW and OFDM radar, respectively; and (b), (d) show code division multiplexing for PMCW and PC-FMCW radar.}
%  \label{fig:schematic_multiplexing} 
%\end{figure*}

Code-division multiplexing (CDM) for MIMO is achieved through the modulation of the radar signal with discrete phase shifts. A typical implementation is the phase-modulated continuous wave radar (PMCW) \cite{bourdoux_pmcw_2016}, where a sequence with favorable autocorrelation properties modulates a carrier signal, and simultaneous MIMO can be implemented through codes with low cross-correlation properties. While low cross-correlation minimizes self-interference, a dynamic range loss occurs that increases with the number of transmitters. Code multiplexing is also used in Phase-coded frequency-modulated continuous wave (PC-FMCW) \cite{petrov2023phase}. It uses a chirp sequence typical to civil radar applications, such as automotive radar, with the addition of an intra-chirp phase modulation sequence. This coding enables simultaneous multiplexing of signals from multiple transmitters.

An alternative approach to multiplexing involves utilizing non-overlapping time and frequency resources for different transmitters. Time-frequency multiplexing typically reduces the maximum unambiguous Doppler or range estimation due to the reduction in sampling rate of slow-time or fast-time, respectively. Non-uniform (random or optimized) sampling in the time-frequency domains can avoid this drawback, albeit at the cost of a diminished dynamic range in range and/or Doppler estimation \cite{knill_random_2019}. Techniques to mitigate this loss in dynamic range, such as sparse reconstruction, have been studied in the literature \cite{correas-serrano_experimental_2018}. Time-frequency multiplexing for MIMO radar is commonly implemented as time-division multiplexing (TDM) in chirp-sequence radars, where different chirps in a sequence are assigned to different transmitters. Non-uniform frequency multiplexing (FDM) can also be easily implemented in OFDM radars \cite{knill_random_2019}, while combined time-frequency multiplexing (TFDM) is also possible in OFDM and other recently proposed waveforms such as orthogonal time-frequency space (OTFS) radar. \cite{correas2023mimo}.

This contribution presents a simulation-based comparison of popular MIMO radar implementations without unambiguity parameter estimation reduction. A comparison is drawn regarding scalability to a high number of transmitters. Furthermore, we briefly discuss practical considerations of the different approaches.

\section{Dynamic range trade-offs in MIMO radar}

\begin{figure} 

    \centering
  \subfloat{%
       \includegraphics[width=0.9\linewidth]{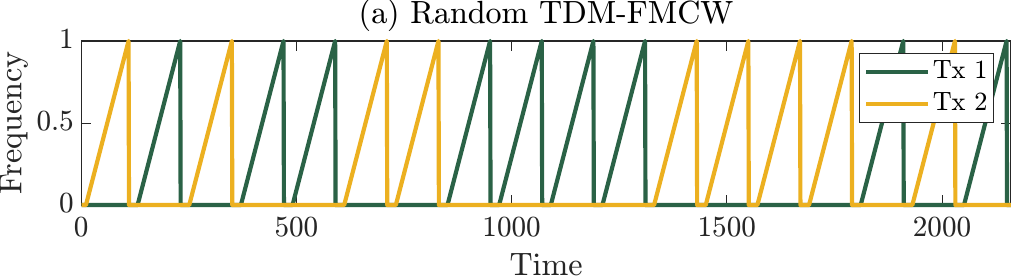}}
	\newline
  \subfloat{%
        \includegraphics[width=0.9\linewidth]{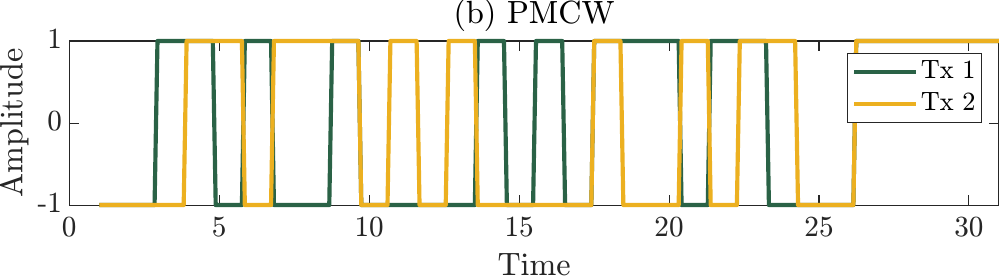}}    
    \newline
  \subfloat{%
        \includegraphics[width=0.95\linewidth]{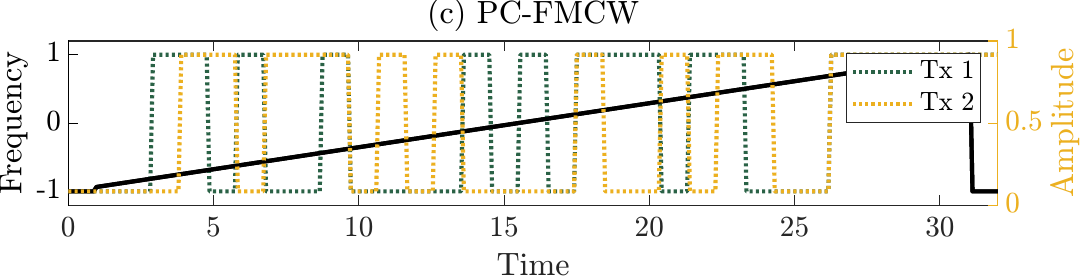}}    
    \newline
    \subfloat{%
       \includegraphics[width=1\linewidth]{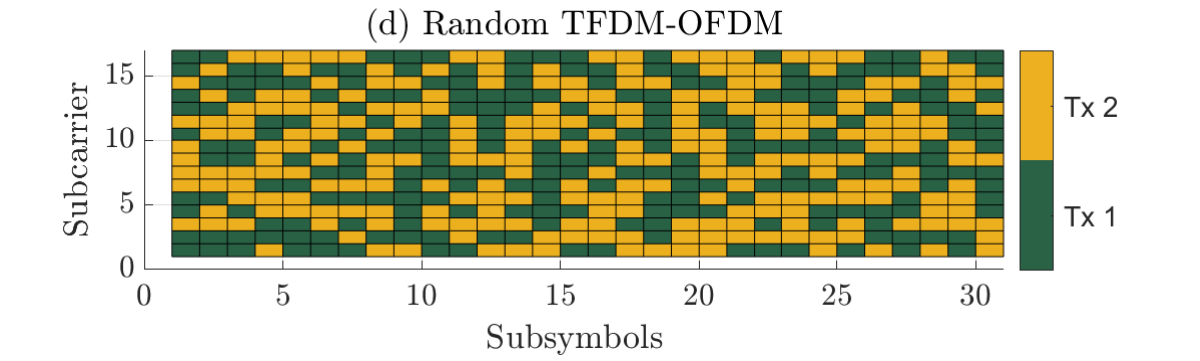}}
\newline
       
  \caption{Schematic depiction of studied waveforms and multiplexing schemes: (a) and (d) show non-uniform time and time-frequency multiplexing for FMCW and OFDM radar, respectively; and (b), (c) show code division multiplexing for PMCW and PC-FMCW radar.}
  \label{fig:schematic_multiplexing} 
\end{figure}

In this section, we briefly introduce a selection of MIMO transmitter multiplexing approaches for various waveforms in which the multiplexing of multiple transmitters incurs a trade-off in dynamic range in distance and/or Doppler dimension. The MIMO implementations studied are code multiplexing for PMCW and PC-FMCW MIMO radar and non-uniform time and frequency multiplexing for FMCW and OFDM radar. A simplified depiction of the waveforms resulting from these transmitter multiplexing techniques is shown in Fig. \ref{fig:schematic_multiplexing}.

\subsection{Random TDM FMCW radar}

One of the most common MIMO waveforms used in civil applications is the TDM-MIMO FMCW radar. In this waveform, transmitters are multiplexed in the slow-time domain of a chirp sequence, with each subsequent chirp being transmitted by the next transmitter in a repeating fashion. This reduction in slow-time sampling frequency results in a lower maximum unambiguous Doppler in the estimation. By reordering this sequence into a non-repeating pseudo-random sequence \cite{mateos2019patent} (see Fig. 1a), a non-uniform subsampling of the slow-time domain takes place, resulting in a reduction of dynamic range in the Doppler estimation, instead of a reduction in maximum unambiguous Doppler. Regarding signal processing, standard beat-frequency and inter-chirp phase shift estimation is performed to extract the time and Doppler shifts, respectively.

\subsection{PMCW radar}

In PMCW radars, a phase code with good autocorrelation properties is applied to a carrier signal and transmitted. MIMO implementation is done through code division multiplexing (CDM), i.e., different codes (often from the same family) are transmitted by different transmitters, as shown in Fig. 1b. The transmitter signal can be written as
\begin{equation}\label{eqn:pmcw_1}
 s(t) = \exp\left(j(2\pi f_c t + \phi(t)\right).
\end{equation}
and the received echo after delay $\tau$ and Doppler shift $f_D$ can be written as
\begin{equation}\label{eqn:pmcw_2}
 r(t) = \alpha s(t-\tau) \exp(j 2\pi f_D t)
\end{equation}
where $\alpha$ is the complex magnitude of the path. For range estimation, the received echoes are sampled in baseband and correlated with the reference code associated with each transmitter. In this process, the range estimation and MIMO transmitter discrimination coincide. Multiple repetitions of these sequences are sent to create slow-time samples for Doppler estimation.

\begin{figure*}[t]
  \centering
  \includegraphics[width=1\linewidth]{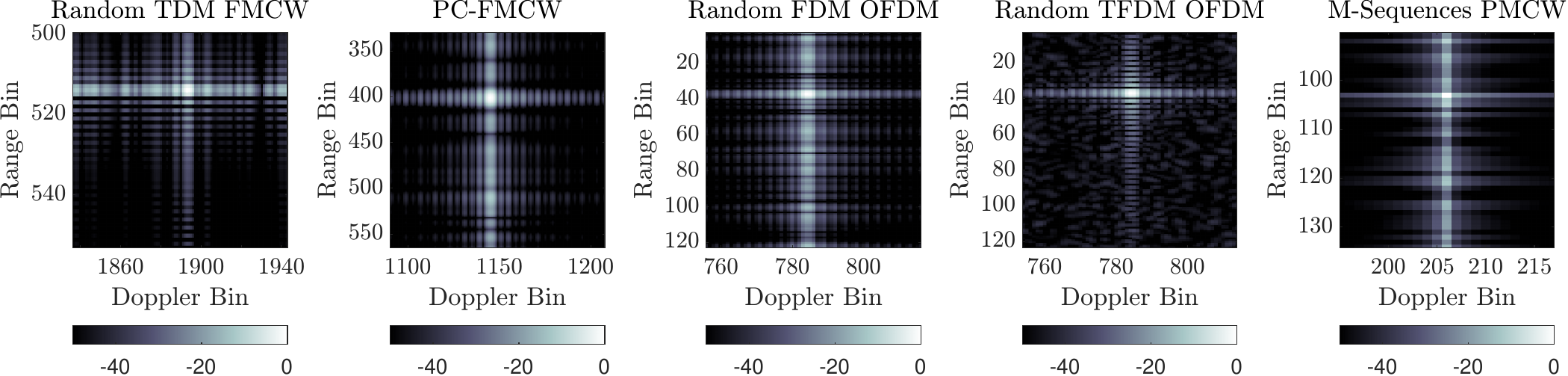}
  \caption[]{Range-Doppler estimation for different MIMO waveforms and $N_\text{Tx} = 12$. Noise-like sidelobes appear in different dimensions of the estimation, depending on the multiplexing approach.}
\label{fig:rd_mimo}
\end{figure*}

\subsection{Phase-coded FMCW radar}

Phase-coded FMCW (PC-FMCW) extends the commonly used FMCW waveform with intra-chirp phase modulation (see Fig. 1c) \cite{kumbul2022smoothed}. The primary purpose of the phase modulation is to enable simultaneous MIMO using code-division multiplexing. Regarding transceiver design, code misalignment must be corrected at the receiver before the signal from different transmitters can be separated. To re-align the codes, a group-delay filter that applies a different delay to each frequency component is typically used,
\begin{equation}\label{eqn:gdf}
 \text{H}_{\text{GDF}}(f) = \exp (j \pi f^2 k^{-1}) 
\end{equation}
\begin{equation}\label{eqn:code_realligned}
 y_{\text{f}}(t,m) = y(t,m)\circledast  \text{h}_{\text{GDF}}(t).
\end{equation}
where $k$ is the chirp modulation rate. For large bandwidth of the codes the group delay filter suffers from signal dispersion that degrades the dynamic range. Several techniques to deal with these effects are discussed in \cite{petrov2023phase}. After code realignment, the decoding by each transmitted code is performed, thereby extracting the signal associated with each transmitter,
\begin{equation}\label{eqn:code_extracted}
 y_d = y_{\text{f}}(t,m)c(t)
\end{equation}
The rest of signal processing consists of standard FMCW beat-frequency and inter-chirp phase shift estimation to extract the range and Doppler radar parameters, respectively.

\subsection{OFDM radar}

At the core of OFDM is the notion of overlapping but orthogonal subcarriers. A single OFDM symbol can be written as
\begin{equation}\label{eqn:ofdm_time}
s(t) = \frac{1}{\sqrt{N_c}} \sum_{n=0}^{N_c - 1} s(n) \exp(j2\pi f_n t)
\end{equation}
where $s(n)$ is the complex amplitude for the $n$-th subcarrier, and $N_c$ is the total number of subcarriers. Multiple OFDM symbols form a typical OFDM radar frame (depicted schematically in Fig. 1d), each preceded by a cyclic prefix (CP) to simplify signal processing. Multiple radar receivers have been studied in the literature \cite{correas-serrano_comparison_2022}, but the spectral-division-based receiver \cite{sturm_spectrally_2013} is the most commonly used in practical applications. This receiver is implemented with an element-wise division of the received signal in the time-frequency domain $\textbf{Y}^{\text{RX}}$ by the time-frequency communication symbols $\textbf{X}$, that is 
\begin{equation}\label{eqn:sm_symbol_2}
 \textbf{Y}^{\text{div}}_{n,m} = \frac{\textbf{Y}^{\text{RX}}_{n,m}}{\textbf{X}_{n,m}}, 
\end{equation}
The range-Doppler map is then extracted by performing an IDFT over the subcarriers and a DFT over the symbols. The multiplexing of multiple transmitters can be implemented by assigning specific subcarriers and/or subsymbols to each transmitter. This paper considers non-uniform allocation in the frequency (random FDM) or time-frequency (random TFDM) dimensions to avoid reducing the maximum unambiguous parameter estimation \cite{correas2023mimo}.

\section{Dynamic range comparison}

In this section, the dynamic range of the MIMO waveforms presented above is studied through simulations. First, Fig. \ref{fig:rd_mimo} shows a section of the noiseless range-Doppler map of the studied MIMO waveforms with 12 transmitters and a single target present. It shows an increase in non-decaying sidelobes (i.e., a reduction in dynamic range) in the Doppler domain for the random TDM-FMCW MIMO waveform and in the range domain for PC-FMCW, Random FDM-MIMO OFDM, and PMCW waveforms. This is consistent with whether the code multiplexing or random sub-sampling happens in the slow or fast-time dimension, which is linked with Doppler and range estimation, respectively. For random TFDM-MIMO OFDM, decaying sidelobes in both range and Doppler around the target can be seen, with lower non-decaying sidelobes spreading over the entire range-Doppler spectrum, indicating an overall lower peak sidelobe, as the sidelobe energy is spread over two dimensions.

Next, a parametric study of the dynamic range of systems with an increased number of transmitters in single-target scenarios is performed. For PMCW and PC-FMCW, an m-sequence of length $M=1023$ is chosen to modulate the signal. For this length, exactly $60$ m-sequences exist, which is set as the maximum number of transmitters. Random patterns are selected for transmitter multiplexing in TDM-FMCW and OFDM radars. The integrated sidelobe ratio (ISLR, see, e.g., \cite{chatzitheodoridi_mismatched_2020} for a definition) is the metric chosen for evaluating the dynamic range. With this metric, we can measure the overall increase in the noise-like sidelobes without needing optimization techniques for peak-sidelobe minimization.

\begin{figure}[t]
  \centering
  \includegraphics[width=0.9\linewidth]{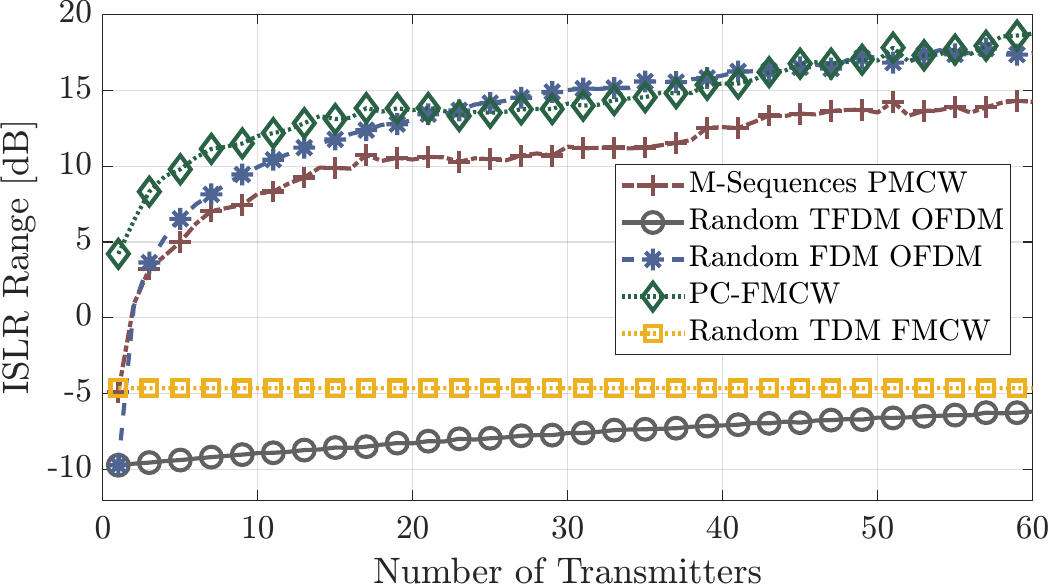}
  \caption[]{Single-channel range cut ISLR for a MIMO system with increasing transmitters..}
\label{fig:islr_range}
\end{figure}

\begin{figure}[t]
  \centering
  \includegraphics[width=0.9\linewidth]{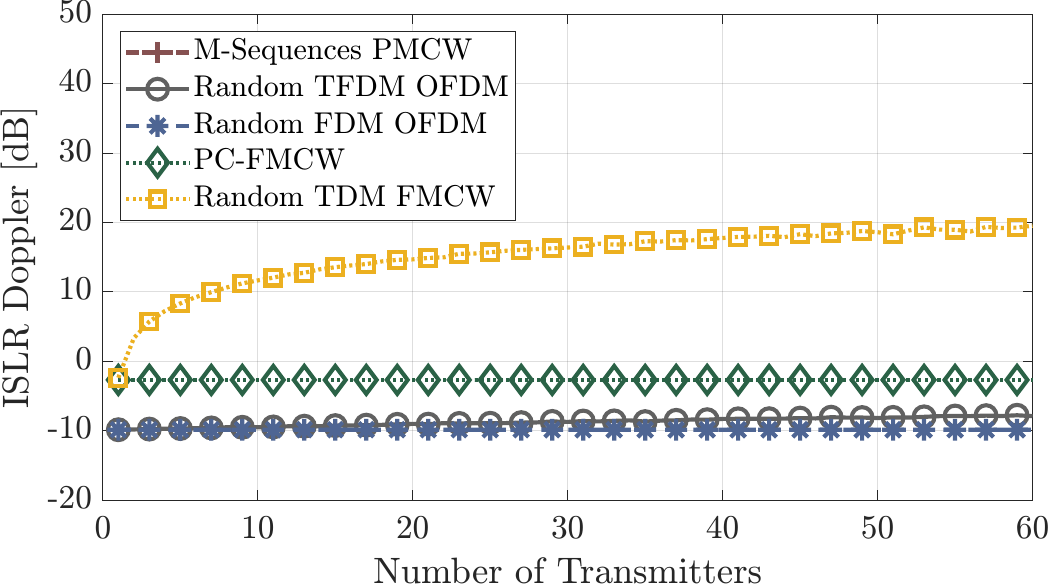}
  \caption[]{Single-channel Doppler cut ISLR for a MIMO system with increasing transmitters.}
\label{fig:islr_doppler}
\end{figure}

The ISLR in range-Doppler is estimated for noiseless scenarios with a single target present. The ISLR in the range and Doppler cuts are shown in Fig. \ref{fig:islr_range} and Fig. \ref{fig:islr_doppler}, respectively. In the range cut, all waveforms that have increased sidelobes in the range domain (namely, PMCW, FDM-MIMO OFDM, and PC-FMCW) show a rapid increase in ISLR when the number of transmitters is low, and it stabilizes as the number of transmitters increases. The ISLR of TFDM-MIMO OFDM increases at a slower rate due to the multiplexing occurring in two dimensions, and the range ISLR of TDM-FMCW remains constant as the number of transmitters increases, as no sparse sampling or code multiplexing occurs in the fast-time dimension. For the Doppler cut, a complementary behavior can be observed. The ISLR increases rapidly in random TDM-FMCW, whereas it remains constant in PMCW, FDM-MIMO OFDM, and PC-FMCW. TFDM-OFDM shows an almost negligible degradation in ISLR as the number of transmitters increases.

\begin{figure}[t]
  \centering
  \includegraphics[width=0.9\linewidth]{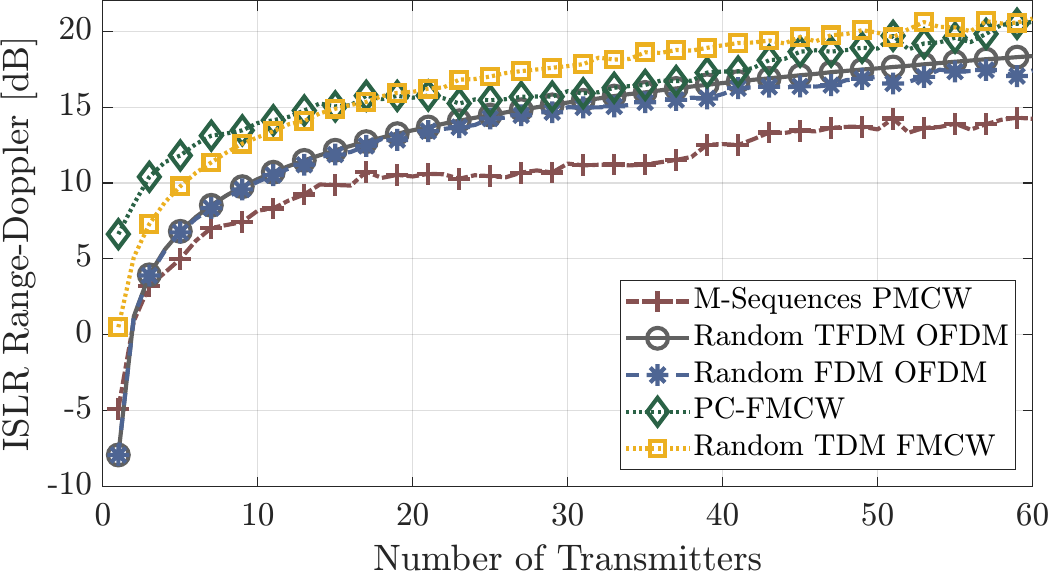}
  \caption[]{Single-channel range-Doppler ISLR for a MIMO system with increasing transmitters.}
\label{fig:islr_range_doppler}
\end{figure}

The ISLR in the entire range-Doppler spectrum is shown in Fig. \ref{fig:islr_range_doppler}, showing the overall sidelobe power in the whole spectrum associated with each MIMO waveform. This value is similar to the range ISLR in the case of PMCW, FDM-MIMO OFDM, and PC-FMCW, as it is dominated by the rising sidelobes in the range cut of the target and similar to the Doppler ISLR in the case of random TDM-MIMO FMCW, as it is dominated by the sidelobes in the Doppler domain caused by the sparse slow-time sampling. In the case of TFDM-MIMO OFDM, however, the ISLR is now comparable to the rest of the waveforms, as the rising sidelobes outside of the target range and Doppler cut are now accounted for. PMCW shows the lowest sidelobes, followed in increasing order by OFDM, PC-FMCW, and TDM-FMCW. 

From an applications perspective, the differences between the dynamic range of the different MIMO approaches have interesting implications for system design. Increased non-decaying sidelobes in the range domain (FDM-MIMO OFDM, PMCW, and PC-FMCW) can be problematic due to near high-power targets masking far low-powered targets that coincide in the same Doppler bin, requiring the use of non-linear sidelobe reducing processing techniques such as sparse reconstruction algorithms. On the other hand, An increase in Doppler sidelobes (random TDM-MIMO FMCW), while generally less disadvantageous in most applications, may make the discrimination of equidistant targets with very different RCS (a common risk scenario in automotive applications and moving target detection against static clutter in surveillance radars) more difficult in the Doppler domain. Finally, the case of TFDM-MIMO OFDM is interesting, as every target present in the scene is masking one another, but with a lower sidelobe level in each of them. This characteristic can be advantageous in applications where the scene is relatively sparse or the noise level is generally above the expected sidelobes.

\section{Conclusions}

This contribution has investigated the scalability of multiple MIMO waveforms that aim to avoid a reduction in maximum unambiguous parameter estimation at the cost of decreased dynamic range. A simulation-based comparison of the performance of the selected waveforms (namely, PC-FMCW, FDM/TFDM-MIMO OFDM, PMCW, and random TDM-MIMO FMCW) shows that the dynamic range degradation affects different parameters in the radar estimation, which can be crucial for application-conscious system design. Moreover, the sidelobe energy present during the estimation of range and Doppler parameters has been evaluated through the ISLR metric for each waveform, revealing insights into how the sidelobe level associated with estimating different parameters increases with the number of transmitters. These results help guide MIMO system design for specific radar applications.

\bibliography{mybib_v3}
\bibliographystyle{IEEEtran}

\vfill

\end{document}